\documentclass[a4paper, times, 10pt,twocolumn]{article}
\usepackage[top=4.9cm,bottom=3.7cm,left=1.5cm,right=1.5cm]{geometry}
\usepackage{ICMLC}
\usepackage{times}
\usepackage{graphicx}
\usepackage{array}
\usepackage{indentfirst}
\usepackage{latexsym}
\usepackage{algorithm}
\usepackage{algorithmic}
\usepackage[font=small]{subfig}
\usepackage{color}
\usepackage{tikz}
\usepackage{booktabs}
\usepackage[margin=8pt,font=footnotesize,labelfont=bf,labelsep=period
]{caption}

\pdfpagewidth=\paperwidth
\pdfpageheight=\paperheight
{%
   \begin{list}{$\bullet$\ \ }
   {%
      \setlength{\itemindent}{0pt}
      \setlength{\leftmargin}{2zw}
       \setlength{\rightmargin}{0zw}
       \setlength{\labelsep}{0.2cm}
       \setlength{\labelwidth}{3zw}
       \setlength{\itemsep}{0em}
       \setlength{\parsep}{0em}
       \setlength{\listparindent}{0zw}
   }
}{%
   \end{list}%
}
\newcommand{\encetcbig}{\includegraphics[width=4em]{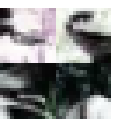}}
\newcommand{\encetcsmall}{\includegraphics[width=4em]{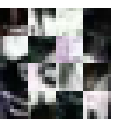}}
\newcommand{\encwetcbig}{\includegraphics[width=4em]{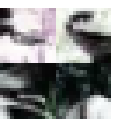}}
\newcommand{\encwetcsmall}{\includegraphics[width=4em]{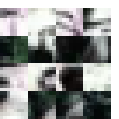}}
\newcommand{\encwetcindbig}{\includegraphics[width=4em]{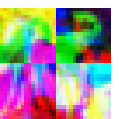}}
\newcommand{\encwetcindsmall}{\includegraphics[width=4em]{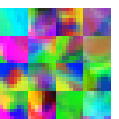}}

\newcommand{\fretcbig}{\includegraphics[width=4em]{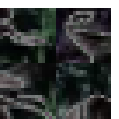}}
\newcommand{\fretcsmall}{\includegraphics[width=4em]{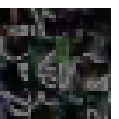}}
\newcommand{\frwetcbig}{\includegraphics[width=4em]{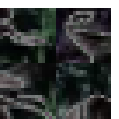}}
\newcommand{\frwetcsmall}{\includegraphics[width=4em]{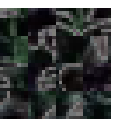}}
\newcommand{\frwetcindbig}{\includegraphics[width=4em]{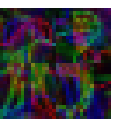}}
\newcommand{\frwetcindsmall}{\includegraphics[width=4em]{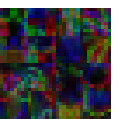}}

\newcommand{\jpsetcbig}{\includegraphics[width=4em]{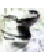}}
\newcommand{\jpsetcsmall}{\includegraphics[width=4em]{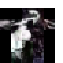}}
\newcommand{\jpswetcsmall}{\includegraphics[width=4em]{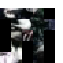}}
\newcommand{\jpswetcindbig}{\includegraphics[width=4em]{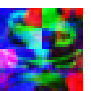}}
\newcommand{\jpswetcindsmall}{\includegraphics[width=4em]{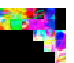}}

\newcolumntype{C}{>{\centering\arraybackslash}m{5em}}
\begin{document}

\title{SECURITY EVALUATION OF COMPRESSIBLE IMAGE ENCRYPTION FOR PRIVACY-PRESERVING IMAGE CLASSIFICATION AGAINST CIPHERTEXT-ONLY ATTACKS}  

\author{\bf{\normalsize{TATSUYA CHUMAN${^1}$, HITOSHI KIYA${^1}$}}\\ 
\\
\normalsize{$^1$Tokyo Metropolitan University, Tokyo, Japan}\\
\normalsize{E-MAIL: chuman-tatsuya1@ed.tmu.ac.jp, kiya@tmu.ac.jp }\\
\\}

\maketitle \thispagestyle{empty}

\begin{abstract}
The security of learnable image encryption schemes for image classification using deep neural networks against several attacks has been discussed.
On the other hand, block scrambling image encryption using the vision transformer has been proposed, which applies to lossless compression methods such as JPEG standard by dividing an image into permuted blocks.
Although robustness of the block scrambling image encryption against jigsaw puzzle solver attacks that utilize a correlation among the blocks has been evaluated under the condition of a large number of encrypted blocks, the security of encrypted images with a small number of blocks has never been evaluated.
In this paper, the security of the block scrambling image encryption against ciphertext-only attacks is evaluated by using jigsaw puzzle solver attacks.
\end{abstract}

\section{Introduction}
In recent years, the rapid development of deep neural networks (DNNs) enables us to implement speech recognition and image classification with high accuracy\cite{lecun2015deeplearning}.
With the rapid growth of DNNs, cloud services like Microsoft Azure and Amazon Web Services are becoming popular instead of using local computing servers owing to the ease of use.
When image classification is performed by cloud computing, security is a concern because the original image must be transmitted to the cloud side.
To protect the security of images for classification under the cloud services, privacy preserving DNNs have been proposed, which encrypt ones before transmission.
Although several image encryption schemes for privacy preserving DNNs have been proposed\cite{tanaka_ICCETW}, \cite{tanaka_AAAI}, \cite{Warit_IEEEtrans}, \cite{AprilPyone_IEEEtrans}, \cite{kiya2022overview}, some attacks successfully restore the visual information of the encrypted images\cite{ito_IEEEaccess}, \cite{Jialong_ICCC}, \cite{warit_gan_2021}.
Therefore, it is essential to use the image encryption method, which can be applied to privacy preserving DNNs and robust against various attacks.
\par
\begin{figure}[t]
	\begin{center}
	\includegraphics[width=\linewidth]{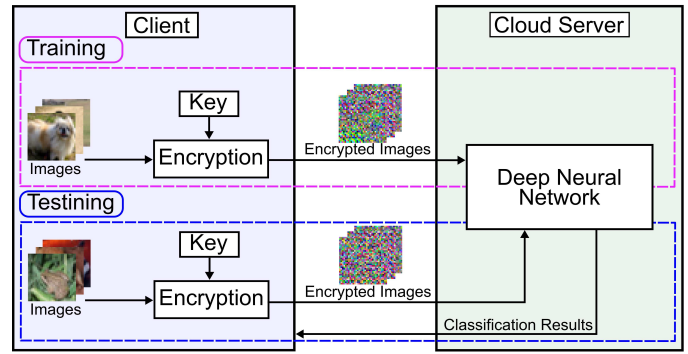}
	\caption{Image classification using deep neural networks}
	\label{fig:scheme}
	\end{center}
\end{figure}
On the other hand, a block scrambling image encryption (BE), which applies to lossless compression methods like JPEG standard, is proposed to generate visually protected images\cite{Kuri_2017}.
The BE is first carried out by dividing an image into blocks; then block-based encryption such as permutation is applied to each block.
In this scheme, $8 \times 8$ or $16 \times 16$-block should be used to avoid decreasing compression performance. 
It has been known that the use of block scrambling enhances robustness against several attacks, but significantly decreases the performance of DNNs.
Owing to the vision transformer (ViT)\cite{dosovitskiy2021an}, a model
for image classification based on the transformer architecture, the BE can be used for privacy-preserving DNNs without a decrease in the performance of DNNs\cite{AprilPyone_compress}.
\par
Robustness of the BE against jigsaw puzzle solver attacks (JPS-attack) that utilize a correlation among the blocks has been evaluated\cite{CHUMAN_IEICE}.
It shows that robustness of the BE against the JPS-attack is enhanced, when encrypted images include a large number of small blocks.
If the resolution of images for image classification is low (like $32 \times 32$ pixels), the total number of blocks in the encrypted image is small.
Therefore, the security of the BE under the condition of a small number of encrypted blocks should be evaluated.
In this paper, the security of the BE against ciphertext-only attacks is evaluated by using the CIFAR-10 dataset which includes color images of $32 \times 32$ pixels.



\section{Preparation}
\subsection{Image Encryption for Deep Learning}
\begin{figure}[!t]
	\captionsetup[subfigure]{justification=centering}
	\centering
	\hspace{1mm}\subfloat[Plain]{\includegraphics[clip, width=2.4cm]{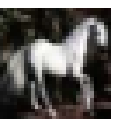}
	\label{fig:orig}}
	\hfil
	\hspace{1mm}\subfloat[BE\cite{Kuri_2017}]{\includegraphics[clip, width=2.4cm]{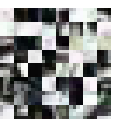}
	\label{fig:etc}}
	\hfil
	\subfloat[EBE]{\includegraphics[clip, width=2.4cm]{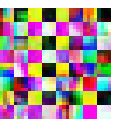}
	\label{fig:wetc}}
	\caption{Example of compressible encrypted images}
	\label{fig:example_enc_comp}
\end{figure}

\begin{figure}[!t]
	\captionsetup[subfigure]{justification=centering}
	\centering
	\hspace{1mm}\subfloat[Tanaka's scheme\cite{tanaka_ICCETW}]{\includegraphics[clip, width=2.4cm]{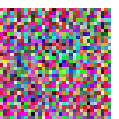}
	\label{fig:le}}
	\hfil
	\hspace{1mm}\subfloat[Madono's scheme\cite{tanaka_AAAI}]{\includegraphics[clip, width=2.4cm]{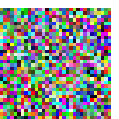}
	\label{fig:ele}}
	\hfil
	\subfloat[Sirichotedumrong's scheme\cite{Warit_IEEEtrans}]{\includegraphics[clip, width=2.4cm]{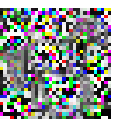}
	\label{fig:pe}}
	\caption{Example of non-compressible encrypted images}
	\label{fig:example_enc}
\end{figure}

Some image encryption schemes have been proposed to apply image classification using deep neural networks as illustrated in Fig.\ref{fig:scheme}.
There are two types of learnable image encryption for deep learning.
One is compressible and other is non-compressible image encryption as illustrated in Fig.\ref{fig:example_enc_comp} and \ref{fig:example_enc}.
Compressible image encryption schemes are carried out by dividing an image into encrypted blocks to apply the JPEG standard.
On the other, non-compressible image encryption schemes are carried out pixel by pixel.
In this paper, we focus on the security of compressible image encryption.



		

\subsection{Block Scrambling Image Encryption}
In a block scrambling image encryption (BE)\cite{Kuri_2017}, an image with $X \times Y$ pixels is first divided into non-overlapped blocks with $B_x \times B_y$.
Next, four steps are applied to the divided image as described below.
\begin{itemize}
\setlength{\leftskip}{0.5cm}
\item[Step 1:] Shuffle the position of each divided block randomly using a random integer which is generated by a secret key $K_1$. Note that all color components in each block are commonly permuted based on $K_1$.
\item[Step 2:] Rotate and flip each block by using a secret key $K_2$, where $K_2$ is commonly used for all color components as step 1. Namely, each divided block is rotated by 0, 90, 180, and 270 degrees and inverted horizontally and vertically according to $K_2$.
\item[Step 3:] Implement negative-positive transformation on each block by using a secret key $K_3$, which is commonly used for all color components.
For instance, a pixel $p$, which is 8-bit pixel values, is transformed to $p'$ by 
\begin{equation}
	\label{negaposi}
p'=
\left\{
\begin{array}{ll}
p & (r(i)=0) \\
p \oplus (2^8-1) & (r(i)=1)
\end{array} ,
\right.
\end{equation}
where $r(i)$ is a random binary integer generated by $K_3$.
In this paper, the value of occurrence probability $P(r(i))$ is fixed to 0.5 to invert bits randomly.
\item[Step 4:] Apply color components shuffling to each block by using a secret key $K_4$.
\end{itemize}
\begin{figure}[t]
	\centering
	\includegraphics[width = \linewidth]{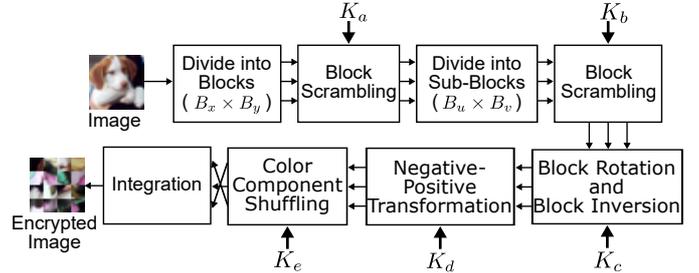}
	\caption{Extended block scrambling image encryption}
	\label{fig:ex_scheme}
\end{figure}
An example of an encrypted image($B_{x} = B_{y} = 4$) is illustrated in Fig.\ref{fig:example_enc_comp}\subref{fig:etc}; Fig.\ref{fig:example_enc_comp}\subref{fig:orig} shows the original one with $32 \times 32$ pixles.

\subsection{Extended Block Scrambling Image Encryption}
To enhance robustness against several attacks, an extended block scrambling image encryption (EBE) was proposed.
In this scheme, block scrambling is carried out two times as illustrated in Fig.\ref{fig:ex_scheme}.
The procedure of the extended block scrambling image encryption is described below.
\begin{itemize}
	\setlength{\leftskip}{0.5cm}
	\item[Step 1:] Permute randomly the divided blocks with $B_{x} \times B_{y}$ pixels using a random integer generated by a secret key $K_a$, where $K_a$ is commonly used for all color components.
	\item[Step 2:] Split each divided block to generate sub-block, each with $B_{s} \times B_{t}$ pixels, and permute randomly the sub-blocks within the block using a random integer generated by a secret key $K_b$. 
	\item[Step 3:] Rotate and invert randomly each sub-block by using a random integer generated by a key $K_c$.
	\item[Step 4:] Apply negative-positive transformation to each sub-block by using a random binary integer generated by a key $K_d$ in accordance with Eq.(\ref{negaposi}).
	\item[Step 5:] Shuffle three color components in each sub-block by using an
	integer randomly selected from six integers by a key $K_e$.
\end{itemize}
In this scheme, although the keys $K_a$ and $K_e$ are commonly used for all color components, $K_b$, $K_c$ and $K_d$ can be used independently. 
Figure \ref{fig:example_enc_comp}\subref{fig:wetc} shows the example of image generated by using the EBE ($B_{x} = B_{y} = 8$, $B_{s} = B_{t} = 4$).
\begin{figure}[t]
	\centering
	\includegraphics[width = \linewidth]{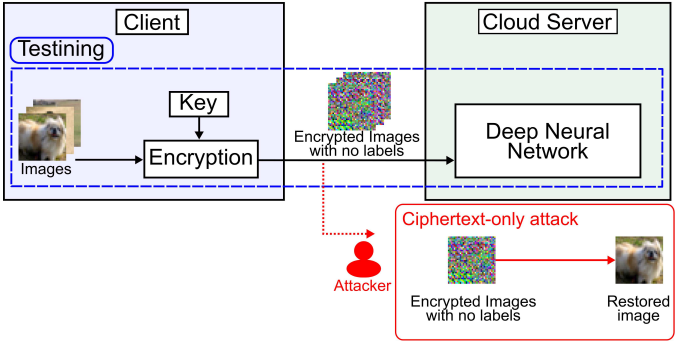}
	\caption{Scenario of restoring visual information from encrypted images with no labels using the ciphertext-only attack}
	\label{fig:senario_test}
\end{figure}
\section{Attacks for Image Encryption}
Several attacks for learnable image encryption have been proposed to restore visual information from encrypted images.
In this paper, we evaluate robustness of encrypted images against ciphertext-only attacks (COAs).
As shown in Fig.\ref{fig:senario_test}, the use of COAs enables us to restore visual information by using the encrypted image without its label.
Brute-force attack, feature reconstruction attack and jigsaw puzzle solver attack are effective to reconstruct encrypted images.
\subsection{Brute-force Attack}
Robustness against brute-force attack is discussed in terms of key space, which depends on the resolution of an encrypted image, encryption steps and its block size.
It has been known that the encryption scheme is robust against the brute-force attack when the key space is larger than $2^{256}$\cite{KURIHARA2015}. 
In the BE, the number of encrypted blocks $n$ is given by
\begin{equation}
	\label{eq:blocknum}
	n = \lfloor \frac{X}{B_x} \rfloor \times \lfloor \frac{Y}{B_y} \rfloor ,
\end{equation}
where $\lfloor \cdot \rfloor$ is a function that rounds down to the nearest integer. 

On the other hand, the number of sub-blocks in the divided block by using the EBE is represented by
\begin{equation}
	\label{eq:blocknum}
	n_{s} = \lfloor \frac{B_x}{B_s} \rfloor \times \lfloor \frac{B_y}{B_t} \rfloor.
\end{equation}

\begin{table}[t]
	\caption{Key space of the BE and EBE}
	\label{table:keyspace}
	\centering
	\begin{tabular}{|c|c|}
	\hline
	Encryption schme & Key space \\ \hline\hline
	BE               &$n! \cdot 8^{n} \cdot 2^{n} \cdot 6^{n} $       \\ \hline
	EBE(common key)              &$n! \cdot n_{s}! \cdot 8^{n_{s}} \cdot 2^{n_{s}} \cdot 6^{n_s} $      \\ \hline
	EBE(independent key)              &$n! \cdot 3n_{s}! \cdot 8^{3n_{s}} \cdot 2^{3n_{s}} $      \\ \hline
	\end{tabular}
\end{table}

Table \ref{table:keyspace} shows the key space of the BE and EBE.
In the BE and EBE, the use of small block size enhances robustness against brute-force attack because the key space increases with the number of divided blocks.

\subsection{Feature Reconstruction Attack}
A feature reconstruction attack (FR-Attack), which uses the edge information on images, is proposed to restore visual information of encrypted ones as described in Algorithm \ref{alg1:alg1}\cite{fra2020}.
It has been known that permuting pixels or blocks in the encrypted image enhances robustness against the FR-attack.
\begin{figure}[h]
	\begin{algorithm}[H]
		\caption{FR-attack\cite{fra2020}}
		\label{alg1:alg1}
		  \begin{algorithmic}[1]
		  \REQUIRE{Encrypted input image $I_{e}$ of size $X \times Y$;\\ \hspace{0.5cm} number of bits $L$; leading bit $b \in \{0,1\} $}
		   \FOR {$p = (u,v) \in I_{e}$}
		   \FOR {$c \in R,G,B$}
		   \IF { $\lfloor p_{c}/(2^{L} - 1)\rfloor \neq b$}
		   \STATE {$p_{c} \leftarrow p_{c}\oplus (2^{L}-1) $}
		   \ENDIF
		   \ENDFOR
		   \ENDFOR
		  \end{algorithmic} 
	\end{algorithm}
\end{figure}

\begin{table}[t]
	\caption{Robustness against the FR-attack and JPS-attack}
	\label{table:result}
	\scalebox{0.9}{
	\begin{tabular}{c|cccccc|}
	\cline{2-7}
	\multicolumn{1}{l|}{} & \multicolumn{6}{c|}{Encryption scheme} \\ \cline{2-7} 
	\multicolumn{1}{l|}{} & \multicolumn{2}{c|}{BE} & \multicolumn{2}{c|}{EBE} & \multicolumn{2}{c|}{EBE} \\ \hline
	\multicolumn{1}{|c|}{Key} & \multicolumn{2}{c|}{common} & \multicolumn{2}{c|}{common} & \multicolumn{2}{c|}{independent} \\ \hline
	\multicolumn{1}{|c|}{\begin{tabular}[c]{@{}c@{}}Block size\\ $B_{x} = B_{y}$\\ ($B_{s} = B_{t})$\end{tabular}} & \multicolumn{1}{c|}{16} & \multicolumn{1}{c|}{8} & \multicolumn{1}{c|}{\begin{tabular}[c]{@{}c@{}}32\\ (16)\end{tabular}} & \multicolumn{1}{c|}{\begin{tabular}[c]{@{}c@{}}16\\ (8)\end{tabular}} & \multicolumn{1}{c|}{\begin{tabular}[c]{@{}c@{}}32\\ (16)\end{tabular}} & \begin{tabular}[c]{@{}c@{}}16\\ (8)\end{tabular} \\ \hline

	\multicolumn{1}{|c|}{Number of blocks} & \multicolumn{1}{c|}{4} & \multicolumn{1}{c|}{16} & \multicolumn{1}{c|}{4} & \multicolumn{1}{c|}{16} & \multicolumn{1}{c|}{4} & 16 \\ \hline
	
	\multicolumn{1}{|c|}{\begin{tabular}[c]{@{}c@{}}FR-Attack\\ (SSIM value)\end{tabular}} & \multicolumn{1}{c|}{0.05} & \multicolumn{1}{c|}{0.04} & \multicolumn{1}{c|}{0.05} & \multicolumn{1}{c|}{0.04} & \multicolumn{1}{c|}{0.06} & 0.05 \\ \hline
	\multicolumn{1}{|c|}{\begin{tabular}[c]{@{}c@{}}JPS-Attack\\ ($L_{c}$ value)\end{tabular}} & \multicolumn{1}{c|}{0.94} & \multicolumn{1}{c|}{0.51} & \multicolumn{1}{c|}{0.94} & \multicolumn{1}{c|}{0.56} & \multicolumn{1}{c|}{0.25} & 0.06 \\ \hline
	\end{tabular}
	}
	\end{table}

\begin{table*}[t]
	\centering
	\caption{Example of encrypted and restored images by using the FR-attack and JPS-attack}
	\label{table:ex_result}
	\begin{tabular}{l*6{C}@{}}
	\toprule
	Encryption scheme & \multicolumn{2}{c}{BE} & \multicolumn{2}{c}{EBE}  & \multicolumn{2}{c}{EBE}\\
	\midrule
	\multicolumn{1}{c}{Key} & \multicolumn{2}{c}{common} & \multicolumn{2}{c}{common}  & \multicolumn{2}{c}{independent}\\
	\midrule
	\multicolumn{1}{c}{\begin{tabular}[c]{@{}c@{}}Block size\\ $B_x = B_y$\\($B_s = B_t$)\end{tabular}} & \multicolumn{1}{c}{16} &  \multicolumn{1}{c}{8} &\multicolumn{1}{c}{\begin{tabular}[c]{@{}c@{}}32\\ (16)\end{tabular}} & \multicolumn{1}{c}{\begin{tabular}[c]{@{}c@{}}16\\ (8)\end{tabular}} & \multicolumn{1}{c}{\begin{tabular}[c]{@{}c@{}}32\\ (16)\end{tabular}} & \multicolumn{1}{c}{\begin{tabular}[c]{@{}c@{}}16\\ (8)\end{tabular}}\\ 
	\midrule
	Encrypted image & \encetcbig & \encetcsmall & \encwetcbig & \encwetcsmall & \encwetcindbig & \encwetcindsmall \\ 
	FR-Attack & \fretcbig & \fretcsmall & \frwetcbig & \frwetcsmall & \frwetcindbig & \frwetcindsmall  \\ 
	JPS-Attack & \jpsetcbig & \jpsetcsmall & \jpsetcbig & \jpswetcsmall & \jpswetcindbig & \jpswetcindsmall \\  
	\bottomrule 
	\end{tabular}
\end{table*}

\subsection{Jigsaw Puzzle Solver Attack}
A jigsaw puzzle solver attack (JPS-attack) is performed by regarding the permuted blocks as pieces of a jigsaw puzzle.
Compared to the FR-attack, the JPS-attack is effective of restoring visual information even when encrypted images include permuted blocks.
It is known that the use of a smaller block size and a large number of blocks in the encrypted image makes JPS-attack difficult to restore visual information\cite{Chuman_IEEEtrans}.

\section{Security Evaluation}
\subsection{Experimental Conditions}
In this section, the security of the BE and EBE was discussed by using the FR-attack and JPS-attack.
We employed 30 images randomly chosen from the CIFAR-10 dataset, which includes $32 \times 32$ color images.
Each image was encrypted separately by using the BE and EBE and then the security of the encrypted image was evaluated.
\par
Robustness against the FR-attack was evaluated by using structural similarity (SSIM) values for the original image and restored one.
In experiments, the average SSIM value for the 30 images was calculated.
Since SSIM values decrease dramatically due to the block scrambling, largest component $L_{c} \in[0,1]$, which means the ratio of the correct pairwise adjacencies, was utilized to evaluate the assembled image by using the JPS-attack\cite{Gallagher_2012_CVPR}.
In this measure, the larger value of $L_{c}$ is, the more visual information is restored.
The extended jigsaw puzzle solver\cite{CHUMAN_IEICE} is utilized as JPS-attack.
Robustness against the JPS-attack is evaluated by using 20 different encrypted images, which were generated from one ordinary image by using different keys.
After assembling 20 encrypted images by the JPS-attack, the image which has the highest $L_{c}$ was selected for evaluation.
We performed this procedure for each encrypted image independently and calculated the average $L_{c}$ for the 30 images.


\subsection{Experimental Results}
Table \ref{table:result} shows the result of security evaluation against the FR-attack and JPS-attack.
Although the large block size like $16 \times 16$ was used for encryption, the BE and EBE are robust against FR-attack.
On the other hand, encrypted images with the BE and EBE using the common key are not robust even when the small block size ( $8 \times 8$) is used for encryption as illustrated in Table \ref{table:ex_result}, where Fig.\ref{fig:example_enc_comp}\subref{fig:orig} is the original one.
However, it was confirmed that the EBE using the independently key is robust against the JPS-attack and FR-attack.

\section{Conclusion}
In this paper, we evaluated the security of compressible image encryption for privacy-preserving image classification against ciphertext-only attacks.
In experiments, robustness of block scrambling image encryption (BE) and extended block scrambling image encryption (EBE) was evaluated by using the feature reconstruction attack (FR-Attack) and jigsaw puzzle solver attack (JPS-attack) as ciphertext-only attacks.
Experimental results show that the jigsaw solver can restore visual information from encrypted images with the BE, even when the block size for encryption is small ($8 \times 8$).
On the other hand, it was confirmed that encrypted images using the EBE are robust against several attacks, even when large block size ($16 \times 16$) is utilized.



\bibliographystyle{IEEEbib}
\begin{small}
\bibliography{refs}
\end{small}

\end{document}